\documentclass[12pt]{article}
\usepackage{amsmath,amssymb,mathrsfs,hyperref,slashed,bbm}
\newcommand{\bra}[1]{\ensuremath{\left\langle#1\right|}}
\newcommand{\ket}[1]{\ensuremath{\left|#1\right\rangle}}
\def\({\left(} \def\){\right)}

\begin{document}
\title{\vspace{-1.8in}
{Realistic clocks for a Universe without time}}
\author{\large K.L.H. Bryan${}^{(1)}$,  A.J.M. Medved${}^{(1,2)}$
\\
\vspace{-.5in} \hspace{-1.5in} \vbox{
 \begin{flushleft}
$^{\textrm{\normalsize (1)\ Department of Physics \& Electronics, Rhodes University,
  Grahamstown 6140, South Africa}}$
$^{\textrm{\normalsize (2)\ National Institute for Theoretical Physics (NITheP), Western Cape 7602,
South Africa}}$
\\ \small \hspace{1.07in}
  g08b1231@ru.ac.za,\  j.medved@ru.ac.za
\end{flushleft}
}}
\date{}
\maketitle

  \begin{abstract}
There are a number of problematic features within the current treatment of time in physical theories, including the ``timelessness'' of the Universe as encapsulated by the Wheeler-DeWitt equation. This paper considers one particular investigation into resolving this issue; a conditional probability interpretation that was first proposed by Page and Wooters. Those authors addressed the apparent timelessness by subdividing a faux Universe into two entangled  parts, ``the clock'' and ``the remainder of the Universe'', and then synchronizing the effective dynamics of the two subsystems by way of conditional probabilities. The current treatment focuses on the possibility of using a (somewhat) realistic clock system; namely, a coherent-state description of a damped harmonic oscillator. This clock proves to be consistent with the conditional probability interpretation; in particular, a standard evolution operator is identified with the position of the clock playing the role of time for the rest of the Universe. Restrictions on the damping factor are determined and, perhaps contrary to expectations, the optimal choice of clock is not necessarily one of minimal damping.
 \end{abstract}


 \section{Introduction}

 Although the concept of time in physical theories is usually taken for granted, it soon produces many problems when attempts are made to
  clearly define the phenomenon or provide a reason behind its features. This so-called ``problem of time'' is typically referred to in the singular, but it is a complex issue  which can be separated into at  least three major concerns. The first of these
  is the apparent necessity for an ``arrow of time'', in spite of the contradictions that arise when the arrow is confronted with  the time-reversal invariance of
quantum mechanics.  This preferred direction of time is
normally attributed to  the second law of
  thermodynamics and still awaiting a convincing explanation for its origin.
  The second is the concern over two different treatments
  of time in physical theories: One within the theory of gravity, where time is dynamical, and the other within quantum mechanics, where time is
  absolute and  akin to time in classical physics. For a working theory of quantum gravity, this difference must
  presumably be resolved.
 
Last but certainly not least is the apparent ``timelessness'' of the Universe. This aspect of the problem of time was introduced by
  Wheeler and DeWitt, who formalized the problem into a relation that has become to be known as the Wheeler--DeWitt
  equation \cite{dewitt},
    \begin{equation}
      \label{wdw}
        \hat{H}\ket{\psi}\;=\;0\;.
    \end{equation}
 Here, $\hat{H}$ is the Hamiltonian constraint of general relativity but  elevated to the status of a quantum operator,
  and $\ket{\psi}$ represents the state of the Universe.
  This expression along with the Schr\"{o}dinger equation make it clear that physical states experience no time evolution.
 This may seem to be  a rather strange requirement to impose on the Universe --- that no time shall pass --- but it is the starting point
  for an interpretation of time that is investigated by the current paper.~\footnote{It has also been a major influence on some other interpretations of time; for example, Rovelli's so-called relational time
\cite{r1,r2}.}
This notion of timelessness inspired Page and Wooters  (PW) to develop their own particular description of ``evolution without evolution''
  \cite{PW} (also see \cite{aha}). It may seem contradictory to obtain evolution within such a timeless Universe, but PW
  managed such a description by partitioning a faux Universe into two distinct subsystems. These are to be identified as ``the clock'', with some associated observable serving as a time variable,  and ``the remainder of the Universe''. For the sake of this paper, these two parts will be referred to as the clock $C$ and the (remaining) system $S$
  respectively.

Along with the viability of  dividing  up  the Universe into $C$ and $S$, this interpretation comes with two further conditions
  which are well summarized by Marletto and Vedral in \cite{vedral} (also see \cite{lloyd}).
  One of these is that the former of the two subsystems
  can be viewed not only as a clock but as a ``good'' one. In this interpretation of time, a good clock is one which has a large number of distinguishable
  states and also, to some approximation, shares no interactions with  the rest of the  Universe. The distinguishable-states requirement allows the clock to
 be measured multiple times, while the no-interactions requirement ensures that $C$ and $S$ are not interfering with one another as
  they effectively evolve.
  A mathematical description of the latter can be stated in terms of Hamiltonians and identity operators for the respective subsystems:
    \begin{equation}
      \label{sep}
        \hat{H}\;=\;\hat{H}_C\otimes {\hat I}_{S} \;+\;
{\hat I}_{C}\otimes \hat{H}_S\;,
    \end{equation}
  Although any real-world scenario would naturally include some interactions  between the two subsystems,
  $C$ can always  be chosen to be small enough so that any influence it has on $S$ is negligible, but then the converse
would not be true.  This point is clarified
  in Section 3.

  The remaining condition on PW's interpretation is the requirement of entanglement between $C$ and $S$ (but notice
that this would follow automatically if the Universe is attributed  with being in a pure state). Importantly, it is
this quantum entanglement that essentially ``sources'' the PW
 description of time evolution \cite{PW}.
  With these three conditions in tow,  the PW proposal  can provide a meaningful  interpretation of time, for
  which the  key aspect is a joint description of a measurement of $S$ and a corresponding measurement
  of $C$. The two measurements are formally linked by way of a conditional probability;
  hence, the PW framework sometimes goes by the name of
  the conditional probability interpretation.

It is worthwhile  to elaborate further on the logistics: In standard quantum mechanics,
 the probability of
  $S$ being in a certain state at a certain time
 takes the time variable for granted as time is treated as an absolute. This is much in line with the
  treatment of time
  in Newtonian mechanics. The crucial change
  in the PW interpretation is that  time should  now be viewed as  an
  explicit consequence of a measurement on the clock and, as
such,
can no longer be taken for granted as an absolute quantity.
  Specifically, the probability for a certain measurement
  of $S$ is conditioned on the probability of measuring a certain time which
is, in reality,  a particular measurement of $C$.
  In this way, a specific  moment in  time can always be
  assigned to  any given measurement on $S$.
 
It is important to note that there is an integration variable which inherits the role of classical time within the PW description
  of evolution.
  This integration variable, denoted here by $n$, provides an ``abstract time'' for describing the evolution of $C$. To be clear, $n$  takes the usual place of $t$ in the evolution operator for $C$ and, then vicariously (via the systems'  mutual entanglement), in the evolution operator for $S$.
 However, as $n$ has no real physical significance, it  is necessarily
  integrated out of the description; leaving only the measurement of some clock observable to provide a time parameter for
  $S$. The procedure will be made clearer later in the paper when explicit descriptions of the probabilities are given.
  An alternative approach to the manual inclusion of $n$ is to introduce an ancillary system
  which is eventually traced out as in \cite{lloyd}.

  PW's approach to time has not been without criticism. Most noteworthy is that of  Kuchar, who
  questioned the ability of the conditional probabilities to display  any evolution at all \cite{kuchar}.
  This was essentially a question of whether a dynamical description could be obtained
  using this interpretation or whether such a description was inevitably static and unable to
  describe change.
  These concerns of Kuchar have since been investigated by Dolby \cite{dolby}. By presenting a more explicit definition of the PW conditional
  probabilities, Dolby showed that measurements at successive time intervals
  (and so dynamical descriptions) were in fact possible.
  Further investigation has more recently been  carried out by Giovannetti, Lloyd and Maccone \cite{lloyd}. Through an independent approach,
  Giovannetti \textit{et al} showed that using the conditions of PW's interpretation allowed for a description of measurements
  at successive points in time within the framework of modern quantum-information theory.
  These two responses individually put to rest the concerns of Kuchar; the conditional probability approach
  does indeed allow for  a dynamical description of subsystems, even within the confines of a timeless description
of the Universe.

  Since the PW interpretation requires specifying a clock system, it is of interest to investigate what types of clocks
  are consistent with such a framework. Indeed, this was pursued by Cornish and Corbin (CC) in \cite{CC}, where they built upon the formalism
 that was already developed by Dolby. Whereas Dolby's specific example was a toy model
  with the Hamiltonian for the clock given by $\;H_C=p\;$, CC
  rather employed a free particle as the clock so that the Hamiltonian is given by $\;H_C=p^2/2m\;$ ($p$ and $m$ are the momentum and mass of the clock respectively).  At this point, it
  is worth recalling that the clocks in PW's interpretation do not measure time directly. Instead, another clock variable
  (such as position) is used as a time parameter for the system. For instance, by using classical relations between time,
  position, and momentum, CC translate the position of the particle into a corresponding time measurement.
 The more conventional notion of time remains  an ``un-observable'' in the sense that no operator can be used to directly measure this abstract parameter.
 
  In order to further assess the applicability of the
  conditional probability  framework, an investigation into realistic clocks would be helpful.
  \footnote{In this regard, the free-particle clock, as proposed by CC,  is not a realistic choice as it can only ever be subjected to
one measurement.}
  The choice of clock for the current study
 is a damped harmonic oscillator, as this is a reasonable facsimile  of the types of clocks that are encountered in ``real-world'' scenarios.
The analysis  shows that the evolution of $S$ can be faithfully described in terms of a specific measurement of time,
which is  actually the position
 of the oscillator clock $C$.
This treatment, following that of CC, also considers the accuracy of the clock and the rate at which such a system decoheres.
 In short,  the conditional probability approach to interpreting time holds up  well under the use of a (somewhat) realistic clock.

The rest of the paper is organized as follows:
  The interpretation of a damped harmonic oscillator as a clock is fleshed out is Section 2. Next,
  a description of the evolution of the system $S$ in terms of the clock's position is explored in Section 3. The outcomes are
  assessed in the final section of the main text, while Appendix A provides some additional  mathematical details.

 \section{The Damped Clock}

  \subsection{Choosing the right clock}

    In order to motivate the current choice of clock, the PW criteria for a good clock will
    first be considered. As discussed in the previous section, a good clock implies that $C$ and $S$ are approximately non-interacting systems --- meaning zero interactions between them in an idealized case.
   If a clock is truly ideal,  then the measurement  of some clock observable  would exactly
    match the abstract time variable $n$ at any given instance.
    The goal of this paper, however, is to further assess the use of PW's interpretation of time for  a more realistic scenario. Therefore, the ability of the clock to approximate a real-world system rather than the ideal case
    becomes a crucial consideration in making a choice.
    In this regard, a well-chosen clock should somehow
account for interactions between it and the
system.~\footnote{The inclusion of interactions was also considered by \cite{lloyd}, but the approach of that paper  differs significantly from the current one.}
    Another consideration is the ability of the clock to be subjected to multiple measurements (in other words, the clock should behave like a clock).

In light of all this, a nice place to start is
   with  a semiclassical (or coherent-state) description of a harmonic oscillator.  This choice of clock has the built-in features of  periodicity and stability; meaning that it readily satisfies the requirement of surviving many successive measurements. The effects of interactions can also be included by further generalizing to the case of a damped harmonic oscillator. However,
an  under-damped  oscillator must then be insisted upon; thus ensuring    that the clock  has a sufficiently long ``running time''  before the damping becomes a hindrance.
The mathematical description and implications of the under-damping restriction are dealt with in greater detail
    later in this section as well as in Appendix \ref{app}.

    Ultimately, there are then two considerations to be balanced against one another: First is the goal of having a clock as close to the idealized case
    as possible for the sake of efficient  time keeping  and second is the need for a realistic clock that properly accounts for interactions between it and the system.
    The choice of a damped harmonic oscillator is meant to incorporate the best of both scenarios within one description.

  \subsection{Keeping time}

 As already discussed,
    the clock in the current treatment will measure time indirectly through another variable, which is taken as the position of the oscillator $x$ for concreteness.
    The wavefunction of the clock $\Psi_C$ will then be given by the  coherent-state wavefunction of a damped oscillator in the position representation, which  is expressible as \cite{DHO}
      \begin{equation}
       \label{wf}
       \Psi_C(x,n)\;=\;\langle x|\alpha(n)\rangle\;=\; A e^{-\frac{e^{-rn}m\omega}{2\hbar}(x-\langle x\rangle)^2 + i\phi}\;,
      \end{equation}
    where $A$ is a normalization constant, $\;\langle x\rangle=e^{-rn/2}\sqrt{\frac{2\hbar}{m\omega}}\mathcal{R}(\alpha)cos(\Omega n)\;$
    is the expectation value of position,  $\Omega$/$\omega$ is the frequency with/without damping, $r$ is the damping coefficient, $\alpha$ is the usual coherent-state (complex) parameter and $i\phi$ is an irrelevant phase.
 Notice that the evolution is described in terms of the abstract time  $n$, which will eventually
    be integrated out of the conditional-probability formalism (see Section 3).

    Equation~(\ref{wf}) can be obtained by starting with  the standard Hamiltonian for a damped oscillator and then performing
    a particular canonical transformation \cite{gzyl}. The resulting Hamiltonian  is formally the same as that of a simple
    harmonic oscillator --- and so the usual techniques for deriving the coherent-state wavefunction can be applied ---  but now with a time-dependent mass.
    The position and momentum operators $\hat{x}$, $\hat{p}$ for this system are related through \cite{DHO}
     \begin{equation}
        \label{rel}
          \hat{x}\;=\;e^{-rn/2}\sqrt{\frac{2E}{m_0\omega}+\frac{\hat{p}^2}{\omega^2}}\;,
     \end{equation}
where $E$ is the energy of the oscillator and $m_0$ is its unevolved mass.  This relation illustrates how the time parameter $n$ can be linked to  a measurement of position in this context. Further details can be found in \cite{DHO}.

   The quantity of immediate interest is the probability of the clock measuring a particular position $x$  at
(abstract) time $n$.
    From Eq.~(\ref{wf}), this probability is simply $|\psi_C(x,n)|^2$.
    The accuracy of  such a position measurement  can
    be parametrized by the  width of the relevant Gaussian $\delta$,
      \begin{equation}
      \label{wid}
        \delta\;=\; e^{-rn/2}\sqrt{\frac{\hbar}{2m\omega}}\;.
      \end{equation}

    Another measure of accuracy for this choice of clock would be the rate of change of the width.
    As a damped system, the clock $C$ is expected to
    lose energy and so decohere. This suggests a nonlinear behavior which is not described by the usual  brand of quantum evolution but rather by a Lindblad-like evolution equation.
   In a case such as this, a proposed measure for the decoherence rate
    is given by  \cite{Ng}
    $\;\sigma(n)=\frac{\partial \Big(\delta(n)_{min}^2\Big)}{\partial n}\;$. For the damped harmonic oscillator, this works out as
    \begin{equation}
          \label{ddec}
            \begin{split}
              \sigma(n) \;=\;\frac{r\hbar e^{-rn}}{m\omega}\;.
            \end{split}
    \end{equation}
    This result allows an assessment of how quickly the damped oscillator loses accuracy; in other words, on what
    time scale it decoheres.

   \subsection{Limiting the damping}

  Perhaps contrary to one's expectations, it will be shown that a minimally damped ($r\rightarrow 0$) oscillator is not  necessarily a preferential choice of clock in the current context.
   But first it is  recalled that the frequency of a damped oscillator is given by $\;\Omega=\sqrt{\omega^2-r^2/4}\;$. This expression leads to  the only obvious  restriction
    on $r$, the under-damping condition of $\;\frac{r}{2}<\omega\;$.

    Another relevant consideration is the need for ``resetting'' or ``winding up''   the clock after some  specified time $n_{reset}$ so that the damping factor is not allowed to become too large. As made clear in Appendix \ref{app},
a constraint of $\;r n_{reset}< 1\;$ ensures that there is a linear relation between the clock position and $n$; a relation that turns out to be essential for
the condition probability interpretation to succeed (see Section~3).
This  limit on the  running time,
\begin{equation}
n \;<\; n_{reset}\;\lesssim\; \frac{1}{r}\;,
\label{run}
\end{equation}
 can now  be adopted to impose a second condition on the clock.
    It follows that minimizing  $r$ will achieve the longest possible running time. But, on the contrary,  maximizing $r$ (subject to the under-damping condition) realizes a maximally  accurate  clock.
    That this is the case will be clarified below using the results from the previous section.

    The minimization of the decoherence rate, as  determined from $\;\frac{\partial\sigma(n)}{\partial r} = 0\;$, produces the result
    \begin{equation}
        \label{sigmin}
          rn \;=\; 1\;.
    \end{equation}
Given that the inequality~(\ref{run}) is in play and
that the accuracy of the clock is at a premium, this outcome suggests choosing
    $\;r=\frac{1}{n_{reset}}\;$,  as this is the only way
Eq.~(\ref{sigmin}) could ever be realized.
   
The very same conclusion is reached  when  the width $\delta(n)_{min}$
is   minimized with respect to $r$. In this case, the condition becomes
    \begin{equation}
      \label{widmin}
          \frac{-n}{2}\sqrt{\frac{\hbar}{m\omega}}e^{-rn/2} \;=\;0\;.
    \end{equation}
    The only value of $r$ which would satisfy this equation is $\;r\longrightarrow\infty\;$, which obviously fails the already stipulated conditions.
    The largest possible value for $r$ should then  be considered, which
 leads back to the   $\;r=\frac{1}{n_{reset}}\;$, the same as before.
    This discussion  makes it clear that the desired choice of $r$ is a compromise
    between minimizing uncertainty (large $r$) and maximizing running time (small $r$).   The optimal choice then depends on what one's particular requirements are for a  clock.

  \section{Finding the time for S}

\subsection{The conditional probability perspective}

 The next
    aspect to be investigated is the question  of
whether or not  the damped-oscillator clock $C$  is indeed able to keep track of time for its purifying system $S$.
    This query is best addressed by   phrasing the evolution of $C$ and then of $S$ in terms
of conditional probabilities. What is first required is the probability of the clock being in an eigenstate $\ket{x}$ at
    some reference time  $n'$  given that the clock is in an unevolved coherent state at $\;n=0\;$. Mathematically, this can be described as
    \begin{equation}
    \label{prob}
      P\Big(\Psi_C=\Psi_x;n=n'|\Psi_C=\Psi_{\alpha}(0);n=0\Big)\, \;=\;
|\langle x\ket{\alpha(n')}|^2\;,
    \end{equation}
    where $\Psi_C$, $\Psi_{\alpha}$ and $\Psi_x$ represent the clock state, a coherent state and a state of 
definite position  respectively, and the wavefunction on the right-hand side is that
from  Eq.~(\ref{wf}). 

Now, as will become evident after a reading of Subsection~3.2,   the evolution operator for the system can be expressed
in a form like $e^{i{\hat H}_S T}$, where the system time $T$ can be identified with the clock position $x$ (to some level of approximation).
In view of this and  Eq.~(\ref{prob}), there
is also a conditional probability
    for the system $S$ which is of the form
    \begin{equation}
    \label{prob2}
      P\Big(\Psi_S(T=x);\Psi_C=\Psi_x|\Psi_S=\Psi_{in};\Psi_C=\Psi_{\alpha}(0)\Big)\;=\;\int_{0}^{1/r}dn'|\langle x\ket{\alpha(n')}|^2\;,
    \end{equation}
    where $\Psi_S$ is the state of the system and $\Psi_{in}$ is some preselected initial state for $S$.
    The meaning of Eq.~(\ref{prob2}) is the probability of system $S$ being in state $\Psi_S$ at
    time $\;T=x\;$ when the clock has a position of $x$ given that the clock and system have been  initially synchronized such that
    $\;\Psi_S=\Psi_{in}\;$ when $\;\Psi_C=\Psi_{\alpha}(0)\;$.

    It is worth  noting that the amplitude of the square of the wavefunction for an ideal clock (as discussed in Subsection~2.1) would be given by
    \begin{equation}
    \label{ideal}
      |\langle x\ket{\alpha(n')}|^2\;=\;\delta(n(x)-n')\;,
    \end{equation}
    where $n(x)$ is obtained by inverting the expectation value $\;x(n)=\bra{\Psi_C(n)}\hat{x}\ket{\Psi_C(n)}\;$. Then
    \begin{equation}
    \label{ideal_prob}
    \begin{split}
      P\Big(\Psi_S(T=x);\Psi_C=\Psi_x|\Psi_S=\Psi_{in};\Psi_C=\Psi_{\alpha}^{(0)}\Big)\,&=\,\int_{0}^{1/r}dn'\,\delta(n(x)-n')\\
                                                                                        &=\,1\;,
    \end{split}
    \end{equation}
    provided that $\;0<n(x)<1/r\;$.

  \subsection{Telling time}

What is left is to confirm the previous claim about the evolution operator for the system.
This entails relating the Hamiltonians ${\hat H}_C$ and ${\hat H}_S$ to one another, but some care must be taken.
    Although the constraint $\;{\hat H}\ket{\Psi}=0\;$ is in place, it is a requirement on physical states only.
    The relation between the individual Hamiltonians can therefore be expressed
as
    \begin{equation}
     \label{Hs}
\       H_C \;\dot{\approx}\; -H_S\;,
    \end{equation}
    with the operator symbolism and the associated identity operators now implied, and where $\dot\approx$ represents a weakly
    vanishing constraint as defined in Dirac's constraint formalism \cite{dirac}.

    Equation~(\ref{Hs}) would also generally include an interaction term $H_{int}$ accounting
    for the influence of the clock and the system on one another. However, for current purposes, the clock can be
    considered small enough so that its influence on the
    system can be considered negligible. But the converse
statement  is not applicable, and the  missing interactions would indeed greatly influence the clock. Nonetheless, this picture
is fully consistent with the current analysis, as
    the interaction term can be traced out of the final description, leaving
 $H_C$ with a damping factor
    and $H_S$ (approximately) unaffected.

    The utility  of Eq.~(\ref{Hs}) in this discussion lies in its ability to allow a relationship between the evolution
    operators for $S$ and $C$, which implicitly follows from the entanglement between the two
subsystems.
   Since the position of $C$ is supposed to be used as a time marker for a measurement on $S$,  it
    is important to see if the evolution operator for $S$ can be described in terms of this clock
    variable $x$. To this end, one can start by considering the total (combined) system, whose state can always be
expressed in a Schmidt-decomposed form:
\begin{equation}
|\Psi\rangle\;=\; \sum_{j} |\Psi_{j}\rangle \;=\; \sum_{j}c_{j}|\Psi_j\rangle_C|\Psi_j \rangle_S\;,
\end{equation}
where the $c$'s are numerical coefficients.

Now, given that the clock is evolving in abstract time, this should rather be expressed as
    \begin{equation}
     \label{tot}
      \begin{split}
      |\Psi_j\rangle\;=\; c_j\Big(e^{-iH_Cn}\ket{\Psi_j}_C\Big)\ket{\Psi_j}_S &=\; c_je^{-i(H-H_S)n}\ket{\Psi_j}_C\ket{\Psi_j}_S \\
                                                   &\dot\approx\; c_j\ket{\Psi_j}_C\Big(e^{iH_Sn}\ket{\Psi_j}_S\Big)\;,
      \end{split}
    \end{equation}
for each value of the index $j$.
    This allows the evolution
    operator for $S$ to be identified as $\;U_S(n)=e^{iH_sn}\;$.
However, since $n$ is to be integrated out,
    the goal is to replace this abstract time with the position of
    the clock and then determine whether the resulting expression is an acceptable description of
    evolution.

    To accomplish this last task, the semiclassical relationship between position and time
    can be employed  to relate $n$ to a position measurement of the
    clock. In other words, expectation values  will  be used to determine $n(x)$.
    The expectation value for position is then recalled \cite{DHO}
      \begin{equation}
        \label{exp}
          \langle x\rangle \;=\;Ae^{-rn/2}cos(\Omega(\omega,r)n)\;,
      \end{equation}
    where $A$ represents the amplitude for an oscillator without damping.
    Expanding the cosine component in terms of $\Omega n$ and retaining terms only up to linear order, one finds that
      \begin{equation}
       \label{exp2}
        \langle x\rangle \;=\;Ae^{-rn/2}\;\;\;\rightarrow \;\;\;n(x)=\frac{2}{r}\ln\Big(\frac{A}{x}\Big)\;.
      \end{equation}

A  more careful approach would be to find $n(x)$ by  accurately inverting Eq.~(\ref{exp}).
  However, given that $\;n<n_{reset}<\frac{1}{r}\sim\frac{1}{\omega}\;$ is in play, the higher-order terms produce corrections which are considered small enough to ignore
    in the current treatment (which is meant to illustrate a method and
not provide a rigorous analysis). A fuller motivation for this approximation is given in Appendix \ref{app}.

    The substitution of $n(x)$, as provided by Eq.~(\ref{exp2}), can be used in assessing the evolution
    of system $S$, as  described by the operator $U_S(n)=e^{iH_sn}$. For this discussion, there is no
need to specify an exact form of $H_S$ (see the very end of the section).
To proceed,  one should then consider  Eq.~(\ref{exp2}) with the substitution  $\;y=A-x\;$
and notice that
    $\;x\leq A\;$ due to the decay in time of  $x$, and so $\;y\geq 0\;$.
It follows that
    \begin{equation}
     \label{n}
     \begin{split}
      n(x) &= -\frac{2}{r}\ln\Big(\frac{A-y}{A}\Big)\\
           &= -\frac{2}{r}\ln\Big(1-\frac{y}{A}\Big)\\
           &\approx \frac{2}{r}\Big(\frac{y}{A}+\cdots\Big)\;,
     \end{split}
    \end{equation}
    which provides a form of $n(x)$ that can be inserted into the description of $U_S$.
Along with
    the redefinition $\;\tilde{H}_S=\frac{2}{rA}H_S\;$, the evolution operator now becomes
    \begin{equation}
     \label{evo_op}
     U_S\approx e^{i\tilde{H}_S y}\;.
    \end{equation}

    The form of Eq.~(\ref{evo_op}) is that of a standard evolution operator. The conclusion is therefore
    that the position measurement of the clock does indeed provide a useful measure of time for $S$, as long as $\;n<n_{reset}<\frac{1}{r}\sim\frac{1}{\omega}\;$. This also confirms the claims
that were put forth regarding Eq.~(\ref{prob2}).

    One last point remains to be made. Although $y$ (and so $x$) represents a measurable observable of the clock,
    $x$ is a time parameter from the perspective of the system $S$. This comes down to the fact that, from the viewpoint
    of system $S$, the clock position is not an operator of any sort. This can be motivated by  the (approximate) commutation relation
    $\;[\hat{x},\hat{\mathcal{O}_S}]=0\;$ for all operators
$\;\mathcal{O}_S\in S\;$, which follows from the (approximate)
separability of the Hamiltonian. Then, in terms of the previous Schmidt decomposition,
    \begin{equation}
     \label{expec}
     \begin{split}
      \bra{x}\hat{x}\Big(\ket{\Psi_j}_S\ket{\Psi_j}_C\Big) &= \bra{x}\hat{x}\ket{\Psi_j}_C\ket{\Psi_j}_S\\
                                                       &= x\Psi_{C_j}(x)\ket{\Psi_j}_S\;,
     \end{split}
    \end{equation}
   indicating that $x$ is only a number from the perspective of the system $S$.

  \section{Conclusions}

    The damped harmonic oscillator appears to stand up well as a clock within PW's
    interpretation of time.
    As was shown here, the evolution of a system $S$ is describable in such  a way that  the role
of time is  played  by the position of the clock.

    As far as resolving the problems of time goes, one can hope that an improved understanding  of
   the  conditional probability interpretation might  then  account for  the disconnect between the treatments of time in
    relativity and in quantum mechanics. For instance, a quantum measurement could
 be associated with a time which is  measured by a clock experiencing
    relativistic effects rather than a time which has no dynamics of its own.
    The problem of timelessness in already taken care of in such an interpretation since it is built in as a assumption  from the outset.
    This leaves only the problem of a preferred direction for the arrow of time as dictated by the second law of thermodynamics.

    The logical next step in analyzing PW's interpretation of time would be to use an actual clock
   which is found in the real world. An investigation that uses atomic clocks for this purpose is currently underway \cite{us}.

  \section*{Acknowledgments}
                The research of AJMM received support from an NRF Incentive Funding Grant 85353 and  NRF Competitive Programme Grant 93595.  KLHB is supported by an NRF bursary through  Competitive Programme Grant 93595 and a Henderson Scholarship from Rhodes University.

    \appendix
     \section{Relating n and x}
      \label{app}

        Given the definition  $\;\Omega=\sqrt{\omega^2-\frac{r^2}{4}}\;$,
the  assumption $\;r \sim \omega/2\;$ (as based on the requirement of
under damping
 along with an assumed uniformity of scales) and the condition
that  $\;n<n_{reset}<\frac{1}{r}\;$ (as proposed in the main text), then
the following approximation can be made
        \begin{equation}
         \label{appeq1}
          \Omega n \;\sim rn\; \;\ll 1\;.
        \end{equation}
 The task here is to verify if these inputs are sufficient to attain
a linear relation between abstract time $n$ and clock position $x$.

To obtain  an explicit form for $n(x)$, the first-order approximation of
Eq.~(\ref{exp}) is taken.
After the substitution of Eq.~(\ref{appeq1}), the equation of interest~(\ref{exp}) can be expanded as follows:
        \begin{equation}
         \label{appeq2}
          \begin{split}
           \frac{rn}{2}&=\ln\Big[\frac{A}{x}cos(rn)\Big]\\
                       &=\ln\Big(\frac{A}{x}\Big)+\ln\Big[1-\frac{r^2n^2}{2}+\mathcal{O}(r^4n^4)\Big]\\
                       &=\ln\Big(\frac{A}{x}\Big)-\frac{r^2n^2}{2}+\mathcal{O}(r^4n^4)\;.
          \end{split}
        \end{equation}

It is apparent that the limit $\;\frac{A}{x}\rightarrow 1\;$
        demands that $\;n=0\;$, as it otherwise implies that $\;rn<0\;$.
        Meanwhile, for values of $\;n>0\;$ and $A>x$, then
        \begin{equation}
         \label{appeq3}
          \ln\Big(\frac{A}{x}\Big)\;\approx\;\frac{rn}{2}\;\gg\;\frac{r^2n^2}{2}\;.
        \end{equation}

  It follows that $\;\frac{rn}{2}\approx \ln\Big(\frac{A}{x}\Big)\;$ will be a
        good approximation for a description of $n(x)$ up to  linear order
in $rn$.
        Solving for $n$ and defining $\;y=A-x\;$, one finds that, to first
order,
        \begin{equation}
         \label{appeq4}
          \begin{split}
            n\;=\;\frac{2}{rA} y\;,
          \end{split}
        \end{equation}
        as already claimed in  Eq.~(\ref{n}).

\end{document}